# A Neural Network model of Cultural Evolution

Kingsley J.A. Cox and Paul R. Adams
Department of Neurobiology, SUNY Stony Brook, NY, USA

**Abstract**
It has been proposed (Richerson and Boyd, 2008) that human intelligence is underpinned by a ratchet-like process called Cultural Evolution in which ideas, originated by individuals, can selectively spread by social learning and replace older, less fruitful ones. Useful ideas can thus accumulate beyond the lifetime of individuals. Although both social and individual learning are thought to be achieved by the selective activity-dependent adjustment of synaptic strengths in an artificial neural-net-like manner, there have been relatively few attempts to incorporate neural networks into Cultural Evolution models. This has led to controversy and uncertainty about how cultural traits are created, transformed, transmitted and selected. We have constructed a transparent model of Cultural Evolution based on simple neural networks, and here we show that a population of communicating agents can learn progressively better descriptions of its environment. Specifically we generate input vectors by linearly combining hidden "causes", which agents can learn from in a nonlinear, Hebbian, manner, thus discovering synaptic weights that "unmix" the data to reveal the hidden causes. We previously showed that if agents can communicate their current estimates of hidden causes to other agents, in a selective manner we call "Light", the interacting population can learn unmixing weights under conditions where most noninteracting individuals cannot. Here we show that with "Light" a population can learn progressively better solutions, in a ratchet-like manner. Although our model is highly simplified, this simplicity allows insight into cultural evolution mechanisms that have hitherto been obscure, and provides a stepping stone to more complex, but still relatively transparent, analyses.

## Introduction

Discovery, invention, insight and other forms of creativity are usually achieved by individual humans, building on the achievements of other individual humans in a “cultural ratchet” (Tennie et al., 2009) (Newton, 1960). Cumulative creative learning might underpin human intelligence, defined as the ability to detect hidden structure in apparently random patterns (Hutter, 2007; MacKay, 2003). Recently we have modeled this process using simplified neural networks (Cox and Adams, 2023, 2021) that can communicate with each other, in an attempt to better understand the conditions under which such cultural evolution (Boyd and Richerson, 1985; Mesoudi et al., 2006; Richerson and Boyd, 2008) can occur. In our model world, data are generated in the simplest possible way. Specifically, we assume that input patterns are generated from independently fluctuating hidden signals by linear mixing and presented to a neuron-like agent equipped with plastic synapses, in an “Independent Component Analysis” (“ICA”) setting. Agents learn synaptic weights that mirror the mixing coefficients such that their activities accurately mimic the hidden signals. Previous mathematical work has fully characterized learning by individual agents in this highly simplified ICA setting, including situations in which learning can fail (Cox and Adams, 2014, 2009; Elliott, 2012; Hyvärinen et al., 2004). Even in this extremely simplified case learning can fail for a variety of reasons, including inappropriate starting weights, inadequate preprocessing or inaccurate synaptic plasticity (Cox and Adams, 2014). We then allowed multiple agents to communicate with each other, sharing each others’ tentative solutions, to see whether communication could overcome failures in individual learning (Tomasello et al., 2005). Our model does not attempt to explain the origin of shared intentionality, communication, or joint attention. Rather, it assumes their minimal computational consequence: agents are engaged with a common latent structure and can use one another’s partial solutions as informative signals. We found, rather unsurprisingly, that if agents indiscriminately communicate both correct and incorrect solutions, failures of individual learning produce failures of collective learning (Cox and Adams, 2021). However if agents learn more effectively from other agents that have found correct solutions, collective learning can overcome the limitations of individual learning (Cox and Adams, 2023).

In our model the communicating neural networks are almost absurdly simplified, single neurons with only 2 plastic inputs. Nevertheless this framework still captures the essence of the general difficulty individual agents face in learning intelligent behavior, specifically the ability to discover the hidden causes of seemingly random data, by exploiting the statistical structure of the observations. In the ICA model the clues to the hidden structure are contained in the higher order input correlations created by linear mixing. These higher order correlations drive the core nonlinear Hebbian learning process (Hyvärinen and Oja, 1998; Field, 1987). But even in this simple case learning may completely fail, because of residual pairwise correlations, Hebbian crosstalk, or a combination of both problems (Cox and Adams, 2014). Similar, though more severe, issues bedevil nonlinear ICA, and even more elaborate and realistic situations.

In the present study we examine whether selective collective learning from successful individuals can also lead to ratchet-like improvements in performance. We do this by exploiting the way that learning in the ICA setting depends both on the mixing process and on the statistical distribution of the random hidden sources. In particular, since ICA learning is only possible when the hidden sources fluctuate in a non-Gaussian manner, the learning becomes easier the more the source distributions differ from Gaussianity, as measured by for example source kurtosis. However in the real world what is most easily learned is not necessarily most useful, such that progress - a sort of "cultural evolution" - can occur. In the work we present here, we examine whether agents that have learned an easy solution (high kurtosis source) can collectively switch to a more difficult (low kurtosis source) but more useful solution.
Our model is a highly simplified neural network based model of cultural evolution that illuminates conditions for successful collective learning.

**Methods**

**1.** In previous work we explored an imperfect learning paradigm where there were 2 stable fixed points, the IC, representing a useful solution, and the PC, representing useless learning, and agents learn one of these fixed points. In the current work we have a similar system but in this case we have 'perfect 'learning and instead the 2 stable fixed points correspond 2 different non-gaussian sources. In this way it is possible to focus on how populations of interacting agents behave with respect to learning these possible 'useful 'solutions.
We studied the behaviour of a set of interacting 1-neuron "agents". Each agent had 2 variable-weight inputs, and could learn, by activity-dependent weight adjustment, either unsupervised by Hebbian adjustment triggered solely by current input values or supervised using a standard Delta Rule, comparing the agent's current output in response to the current input vector to the output of a "teacher" agent which is also attempting to learn the correct solution. The goal of learning is to find synaptic weights that permit agents to track one or other of the hidden randomly-fluctuating (but nonGaussian) causes or "sources" (the "ICs") that have been linearly mixed with an orthogonal mixing matrix. Each source is drawn from a different nongaussian distribution, one from a Laplacian (LAP) and the other a Logistic (LOG) distribution. The kurtoses of these distributions are different, LAP with a kurtosis of 3 and LOG a kurtosis of 1.2.
Because the LAP source is more nonGaussian than the LOG source, it is easier to find, however, the LOG source might be more useful.

## 2 Unsupervised learning with ICA

The goal in ICA is to unmix a set of independent unit-variance sources that have been mixed by a mixing matrix. In this paper we use an orthogonal mixing matrix (a rotation matrix which does not introduce second order correlations, thus simplifying the problem). The matrix can be varied by changing the angle θ that the first column of **M** makes with the vector $[1,0]^T$, but throughout this paper $\theta$=45 degrees. While this choice of $\theta$ does not affect the underlying learning dynamics or the relative sizes of the basins of attraction, it creates equal mixing of the 2 sources.

$$\mathrm{M} = \begin{bmatrix} cos\theta & -sin\theta \\ sin\theta & cos\theta \end{bmatrix}$$

$$\mathbf{x} = \mathbf{Ms}$$

where **x** is the observable vector of mixed inputs, **s** is a vector of hidden (unobservable) independent sources, and **M** is the mixing matrix. In all our simulations **M** was orthogonal and there were only two randomly fluctuating sources, one with a Laplacian distribution and the other Logistic. This means that there are two learnable solutions, a synaptic weight vector pointing in the same direction as the column of the mixing matrix associated with the either the LAP or LOG source. For simplicity we only consider one output neuron which tries to find a weight vector that will 'unmix' the sources so that the output neuron tracks one of them.

We use the one unit negentropy-maximizing rule (Hyvärinen and Oja, 1998)

$$\Delta\mathbf{w} = k\, \sigma\, \mathbf{x}\, f(\mathbf{w}^T\mathbf{x}), \text{ normalize } \mathbf{w}$$

where k is a learning rate, **w** is the current weight vector, $f(\mathbf{w}^T\mathbf{x})$ is the output (referred to as y later), and σ is either +1 or -1 depending on the nonlinearity and the nonGaussian source statistics (Cichocki et al., 1997). In our case, with n=2, weight vector normalization after each iteration means that the 'basin' of attraction of the learning dynamics lies on the unit circle (which due to symmetry can be contracted to an hemicircle).

Thus the output neuron sums the inputs weighted by the current weight vector **w** and then passes this total 'activation' through a nonlinearity f(y). In this paper we used a cubic ($y^3$) nonlinearity (which maximises kurtosis) and which requires σ = -1 (Hyvärinen et al., 2004). The current weight vector is then increased by an amount proportional to f(y) and finally the updated weight vector is normalized so that no individual weight can become too large. **M** is constrained to be orthogonal so that there are no pairwise correlations introduced into the input vectors **x** (i.e. the source vectors are only rotated). Even if the original mixing is nonorthogonal this condition can be achieved by suitable preprocessing. The weight vector that enables y to track one of the nonGaussian sources is called the independent component (IC).

We denote unsupervised learning as 'U'.

## 3 **Supervised learning**

For supervised learning, agents use the Delta rule (Widrow and Hoff, 1960). The difference between the agent's output y is compared against a teacher's output $y_T$ (in this case another agent's estimate of the current value of the nonGaussian source). Note that the teacher agents may not have already learned the the correct solution. In this case the output y is the dot product of the input vector and the weight vector and has not been put through a nonlinearity.

$$\Delta\mathbf{w} = k\,\mathbf{x}(y - y_T)$$

where Δ**w** is the update in the weight vector and k is a learning rate. We denote supervised learning as ‘S’.

## 4 Light

We added to the foregoing framework a feature we call 'Light' (Cox and Adams, 2023), so that in the delta rule the learning rate is multiplied by a function of the cosine of the current angle, ϕ, of the teacher's weight vector with the IC, a factor we call 'Light'. There are 2 parameters that can be set to change the amount of Light, a magnitude factor *a* and a steepness factor *b* as follows

$$\text{Light} = a(\cos(\phi))^b$$

In this formulation teachers become more effective the closer they are to the correct solution. Light can be considered as an estimate of the value, or utility, of the current solution. In this paper we will refer to the Light factors a and b as Light (a,b), so for instance Light = $2(\cos(\phi))^3$ would be Light (2,3).

## 5 Interacting and non-interacting agents

Each agent is an independent 2 input and 1 output network as described above. An agent alternates between unsupervised (U) or supervised (S) learning. In this paper we explored how agents behaved both in the presence and absence of Light. A run consisted of T iterations where at each iteration a new input vector was used to change the weights of each agent. Under unsupervised learning the agent learns on its own directly from just the input vector and the learning rule (see one unit learning rule above). If S was chosen then one of the other agents was randomly picked as the 'teacher 'for that iteration and the delta rule was used. At each iteration for each agent U and S was chosen with a 50% probability.

Matlab was used for the simulations.

This set-up contains the essence of ratchet-like cultural evolution: individual discovery that can be copied by others, so **useful** learning by a population of agents can be accelerated.

## Results

Previous mathematical analysis of the 1-neuron ICA model showed that with 2 nonGaussian sources there are 2 weight vectors that can be stable equilibria of the learning dynamics (Elliott, 2012).

### Basin sizes

When starting with 100 non-interacting agents with randomly assigned initial weight vectors between [1,1] and [-1,1] we find that 70% go to the LAP fixed point (FP) [1,1], and 30% to the LOG FP, [-1,1]. This we interpret as reflecting the ratio of the basin sizes. This ratio, 5:3 broadly reflects the ratio of the source kurtoses, which is 5:2 (see Appendix).

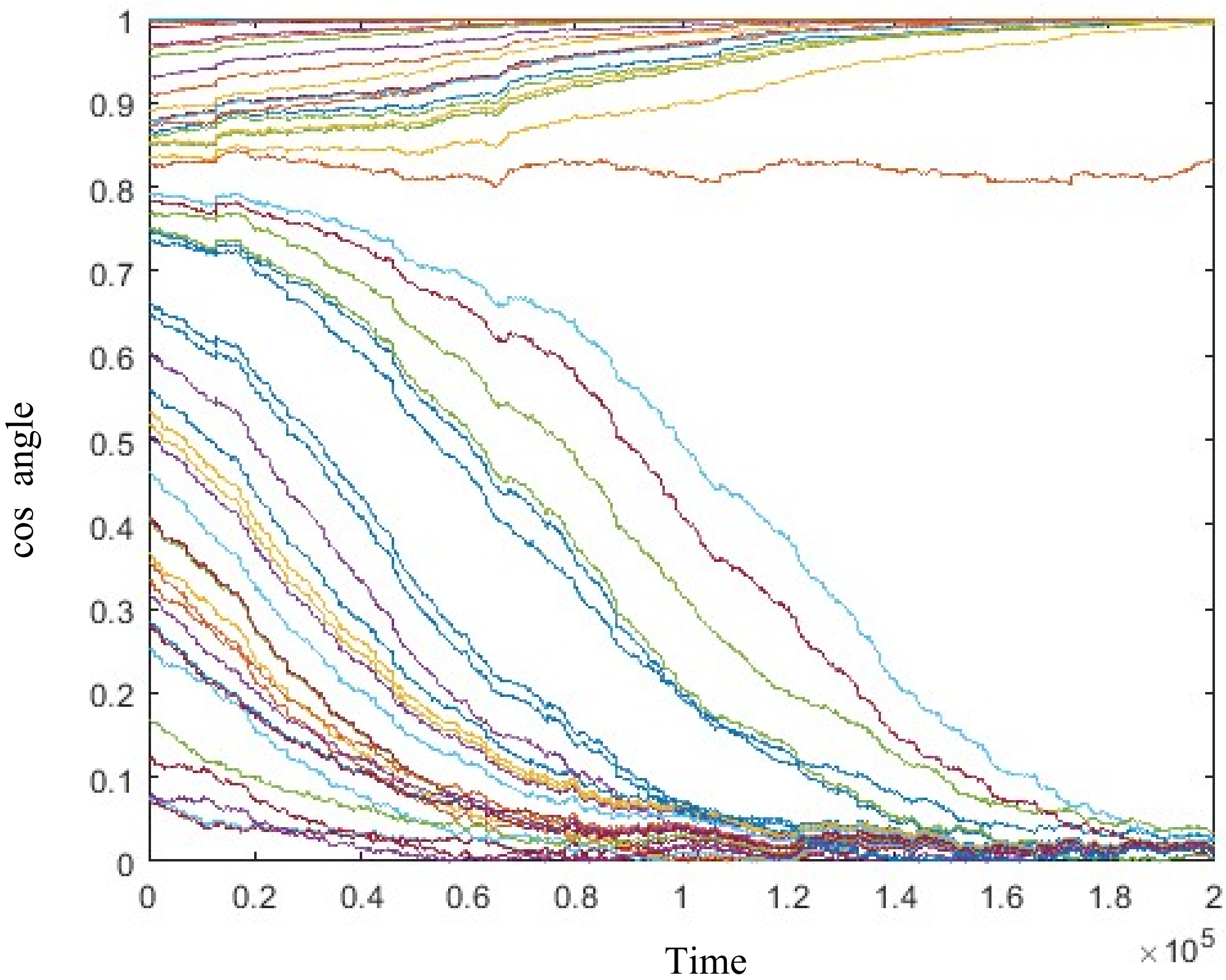

Figure 1 Basins of attraction of the learning dynamics. 100 non-interacting agents were starting at random initial weights to explore the nature of the basins of attraction for the LOG and LAP fixed points. The Y axis is the cosine of the angle each weight vector makes with the LOG IC so when the cosine is 1 then an agent has reached the LOG fixed point. Since the LAP fixed point is orthogonal then when the cosine is 0 then an agent has reached the LAP fixed point. The learning rate was set to be very low (0.0001). One of the agents started off almost exactly on the separatrix and after 200,000 epochs still had not reached a fixed point. Since the cosine is nonlinear, the LAP FP looks noisier that the LOG, but this is just an artefact.

Therefore the source with the higher kurtosis can be viewed as an ʾeasier ʿIC to find than the source with the lower kurtosis. Based on the above we can represent the attractor ʾbasins **of attraction** ʿas a line from -1 to 1 split in a 3:5 LOG:LP ratio as follows with LOG red and LAP blue

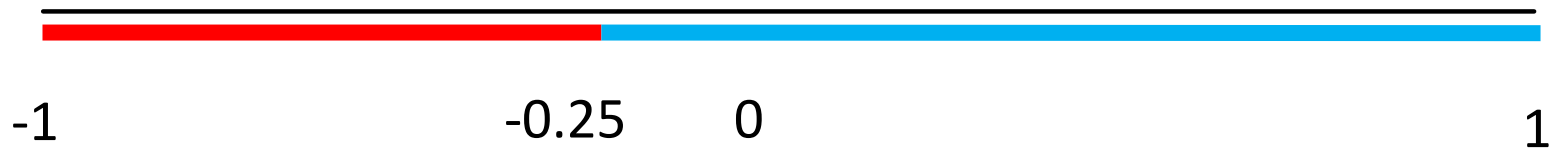

-1 -0.25 0 1

This result shows that an agent will learn either the LAP or LOG source vector depending on what the initial weight vector is, and that the LAP basin is larger than the LOG basin. If the initial weight vector is between [-1,1] and [-0.25,1] then the agent will find the LOG FP, and if the initial weight vector is between [-0.25,1] and [-1,1] then the agent will find the LAP FP. In reality the ‘harder’ solution might actually be more ‘useful’ to the agent.

**Initial explorations with 4 Agents**

**2 agents in LAP basin and 2 agents in LOG basin**

Initially simulations were done with 4 interacting agents, 2 starting at the LAP IC and 2 starting at the LOG IC. Without Light both the LOG agents were pulled over to the LAP fixed point, because supervised learning causes all agents to converge to the same FP (Cox and Adams, 2021). The plot below shows cosines of the angles between the weight vectors of the agents and the LOG FP ([-1,1]).

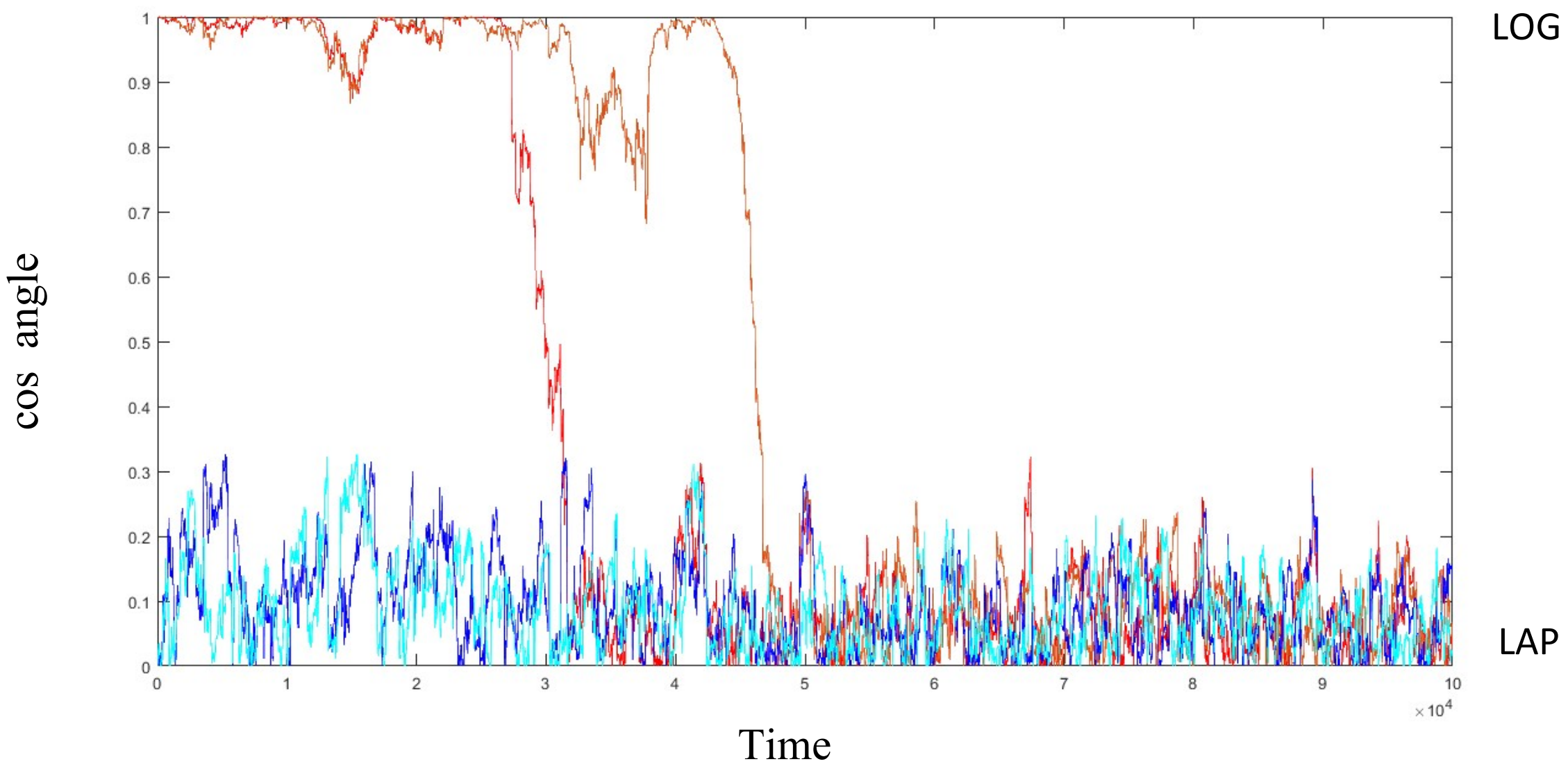

**Figure 2** Agents 1 and 2 (red) starting from LOG IC i.e. [-1,1] and agents 3 and 4 (blue) starting from LAP IC i.e. [1,1]. All agents end up at the higher kurtosis, LAP, IC [1,1]

**With Light**

When Light (see Methods 4) is applied to the LOG agents then both the LAP agents are instead pulled over to the LOG FP. The plot below shows angles between the weight vectors of the agents and the LOG FP ([-1,1]). The plot below shows angles between the weight vectors of the agents and the LOG FP ([1,1]).

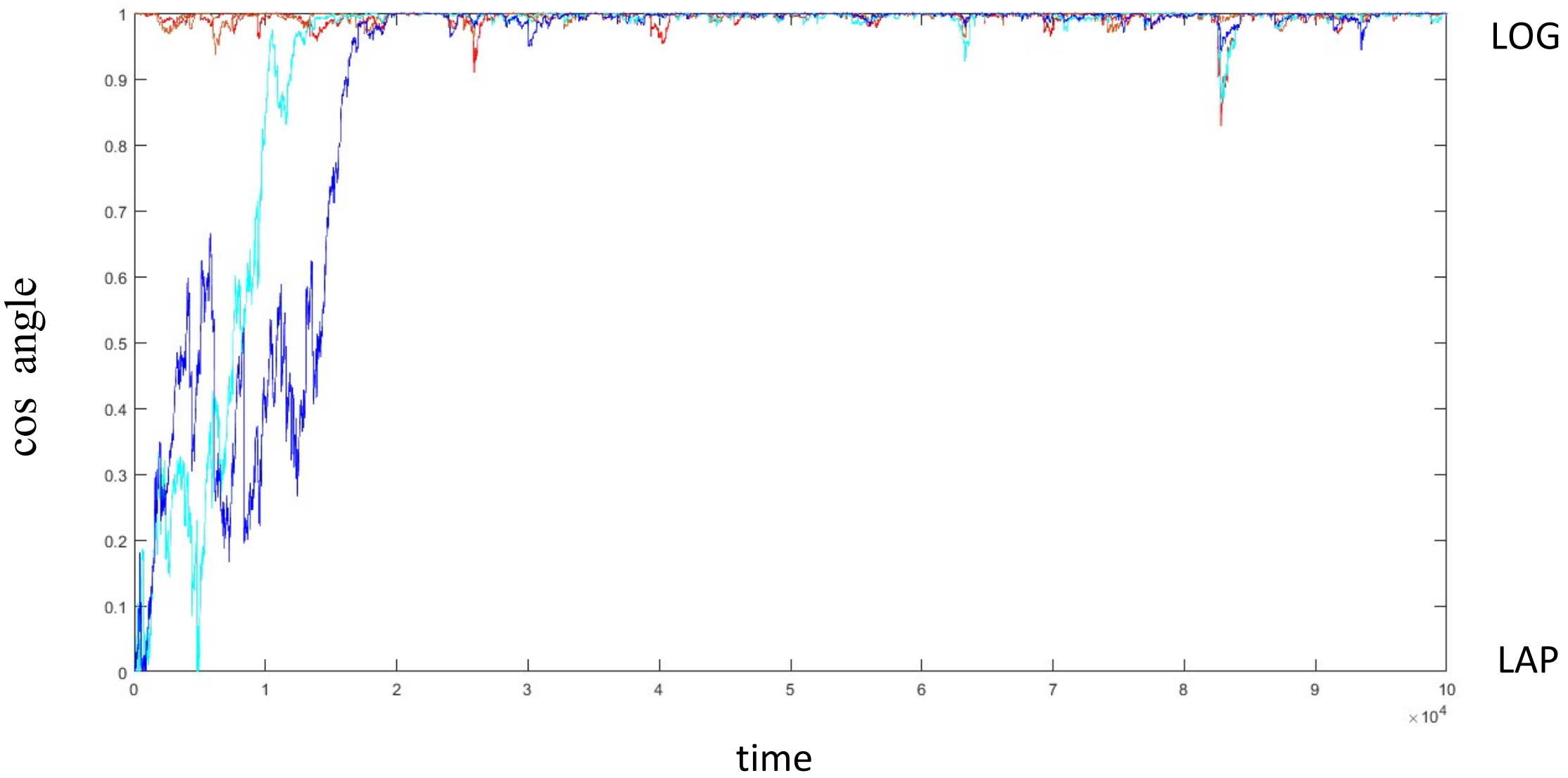

**Figure 3** Agents 1 and 2 (which start at the LOG IC) are red and agents 3 and 4 (starting at the LAP IC) are blue. With Light (2,2) all agents go to the LOG basin [-1,1]

In order to see if just one agent in the smaller basin could attract 3 in the larger basin, 1 agent was set at the LOG IC and the other 3 at the LAP IC, i.e. one agent at [-1,1] and 3 at [1,1]. Under these conditions, without Light **as expected** the LOG agent ended up at the LAP FP.

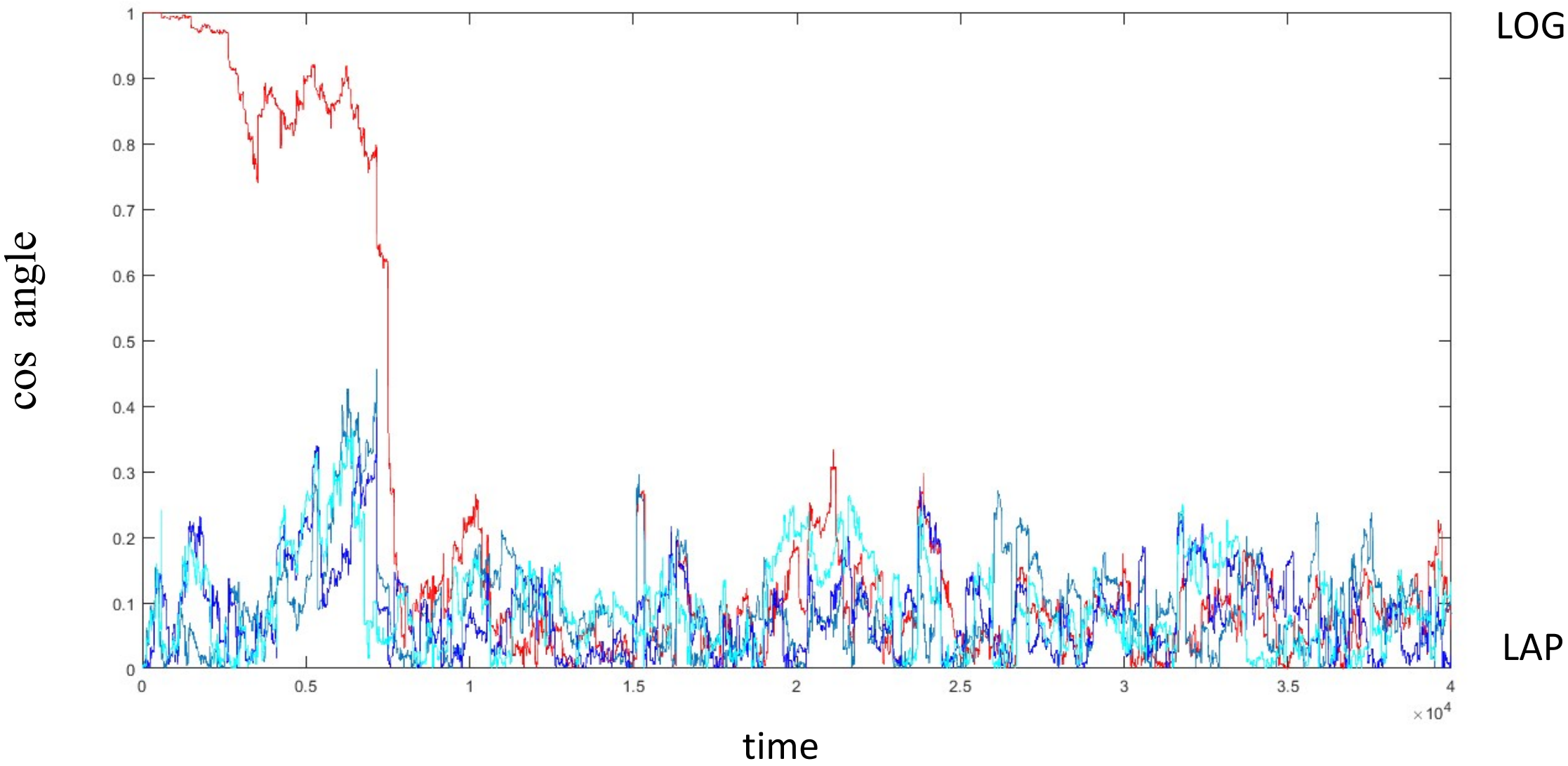

**Figure 4** This plot shows the three LP basin agents (blue) and the LO basin agent (red) all converging to [1,1], i.e. the LAP IC. The three agents that started in the LAP basin pulled the agent in the LOG basin over to the LAP basin.

However, with added Light (2,3) the sole LOG agent **now** pulls the 3 LAP agents into the LOG basin i.e. all go to [-1,1]

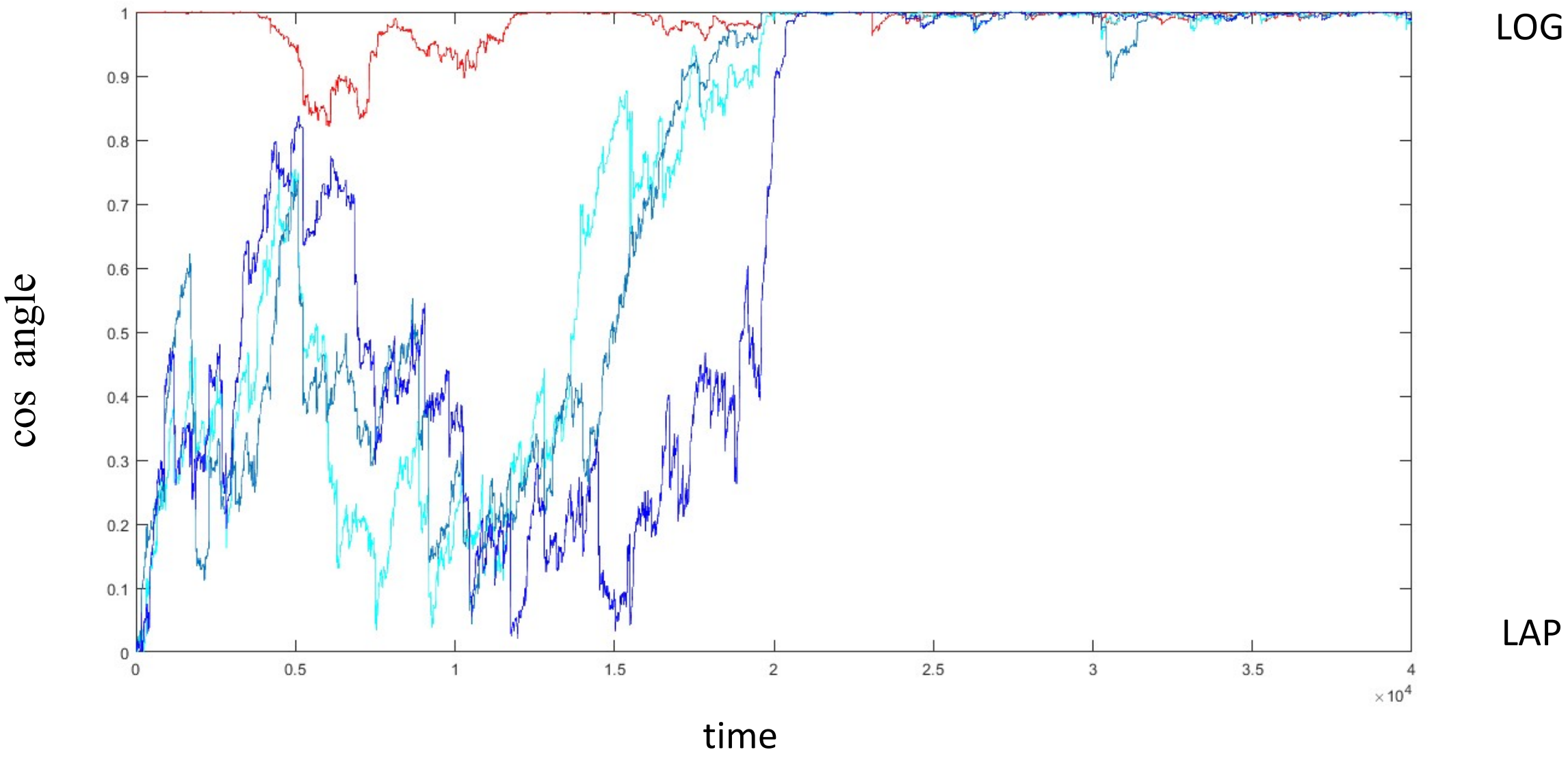

**Figure 5** LOG agent (red) pulling the 3 LAP agents (blue) into the LOG basin with Light (2,3). The LOG agent started at [-1,1] and the 3 LP agents at [1,1].

**N Agents**

The simulations were expanded to explore the behaviour of an arbitrary number of agents. 30 agents are shown below (figure 6) with agents, without Light, starting in the LOG basin initially moving to the LOG IC (1,-1) and agents starting in the LP basin moving to the LAP IC (1,1). Then over time all the LOG agents are pulled over to the LAP IC.

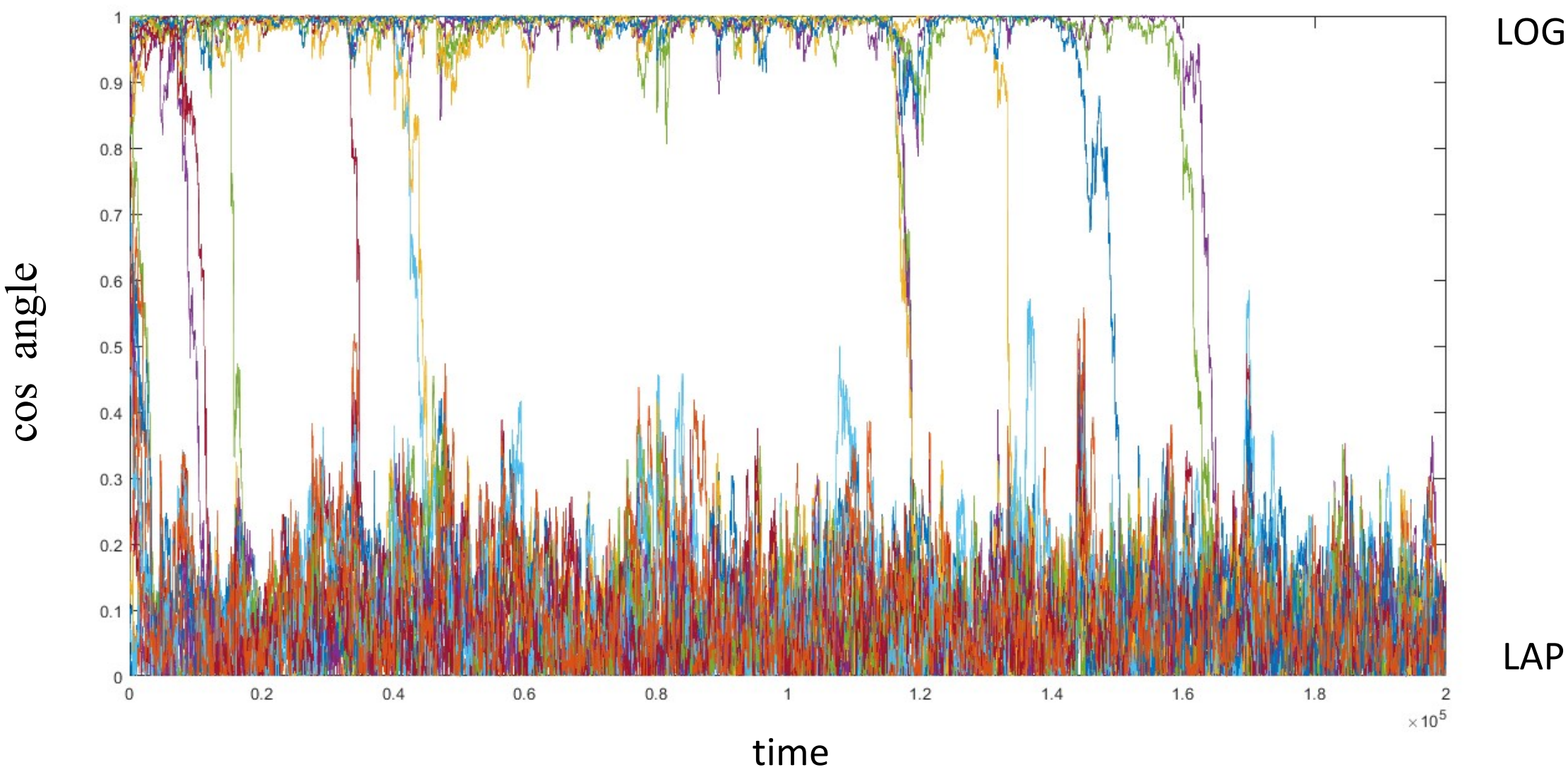

**Figure 6** 30 agents starting at random weight vectors (i.e. extreme left), no Light. The agents that started in the LOG basin quickly went close to the LOG FP (within 50,000 iterations) but over time all were eventually pulled over to the LAP FP

However, with Light (2,3), all the LAP agents are **eventually** pulled into the LOG basin (figure 7). Interestingly, at around 50,000 epochs a LOG agent was pulled to the LAP basin, but later returned.

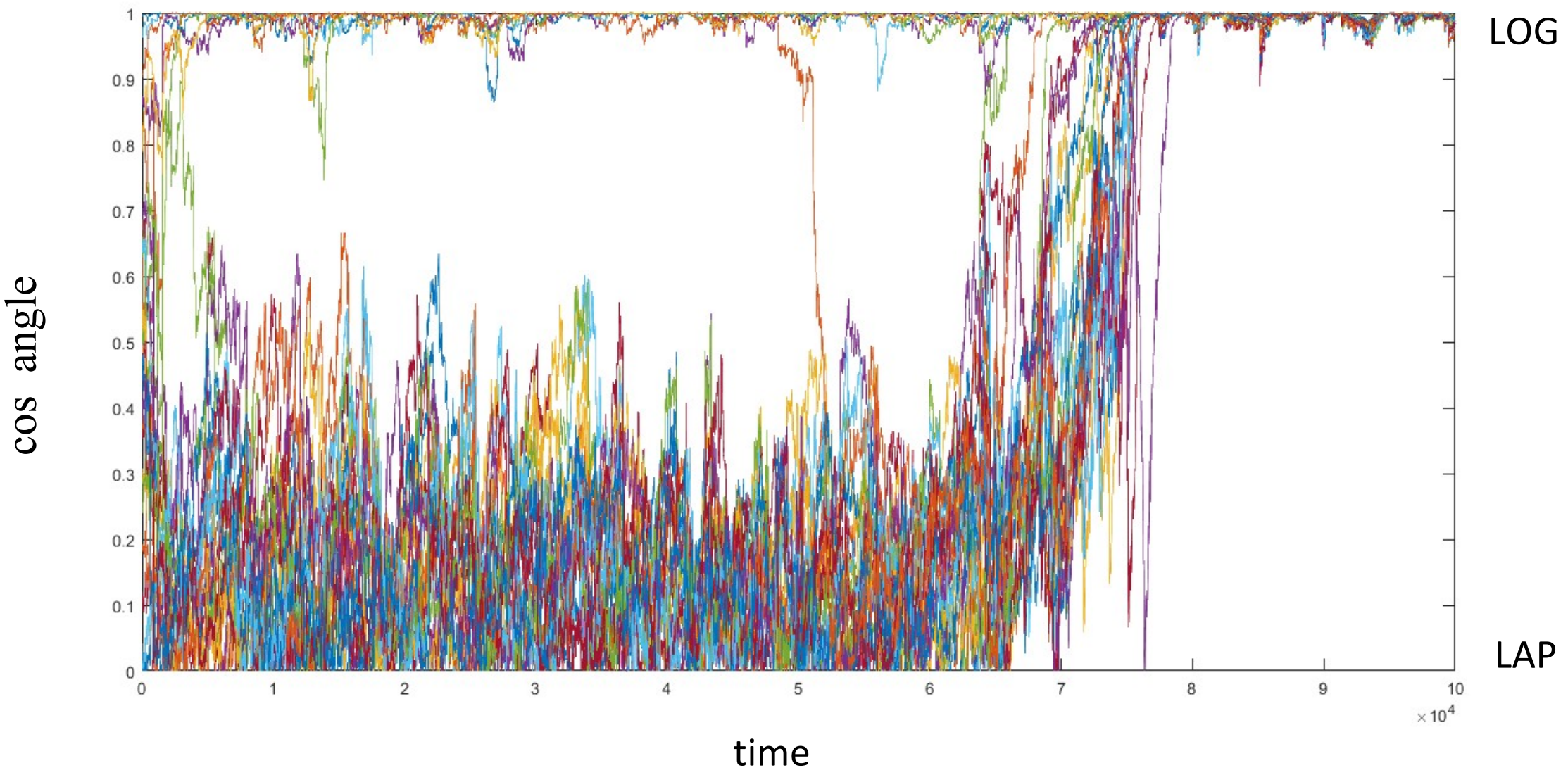

**Figure 7** 30 agents starting at random weight vectors, Light (2,3). Compare with figure 6. In this case, with Light, all the LAP agents are pulled to the LOG FP.

**100 Agents**

We explored further the idea that one agent in one basin can influence many agents in the other basin. We increased the number of agent from 4 (see figure 4 and 5) to 100. In figure 8 one agent was started at the LOG FP and the other 99 at the LAP FP. Without Light unsurprisingly the LOG agent was pulled over to the LAP FP.

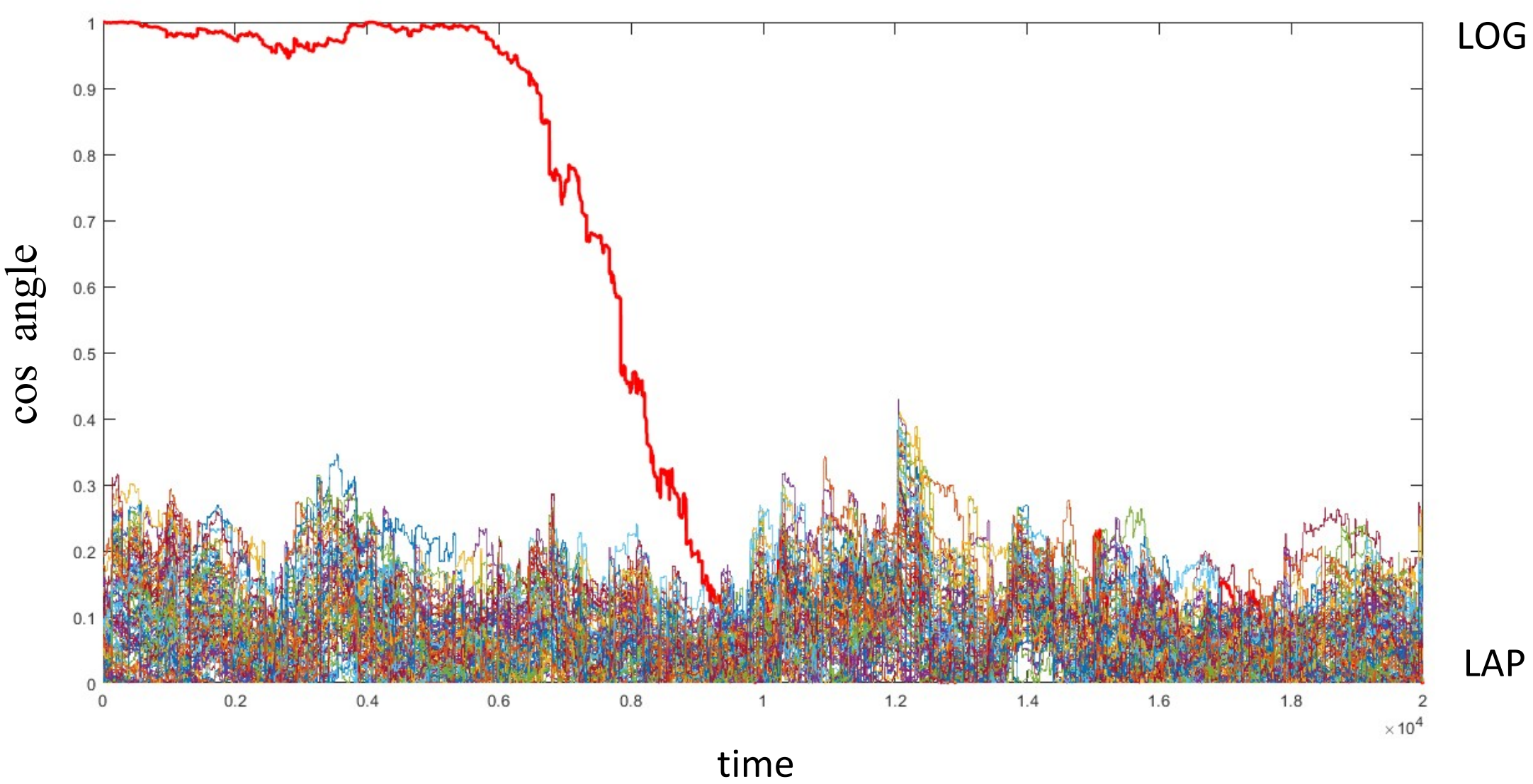

**Figure 8** 100 agents with one starting from (1,-1) (red line), i.e. LOG IC without Light. The LOG agent is pulled over the the LAP FP

With strong enough Light rather surprisingly one LOG agent was able to pull over all 99 LAP agents, even though each LAP agent was looking at another LAP agent 99% of the time.

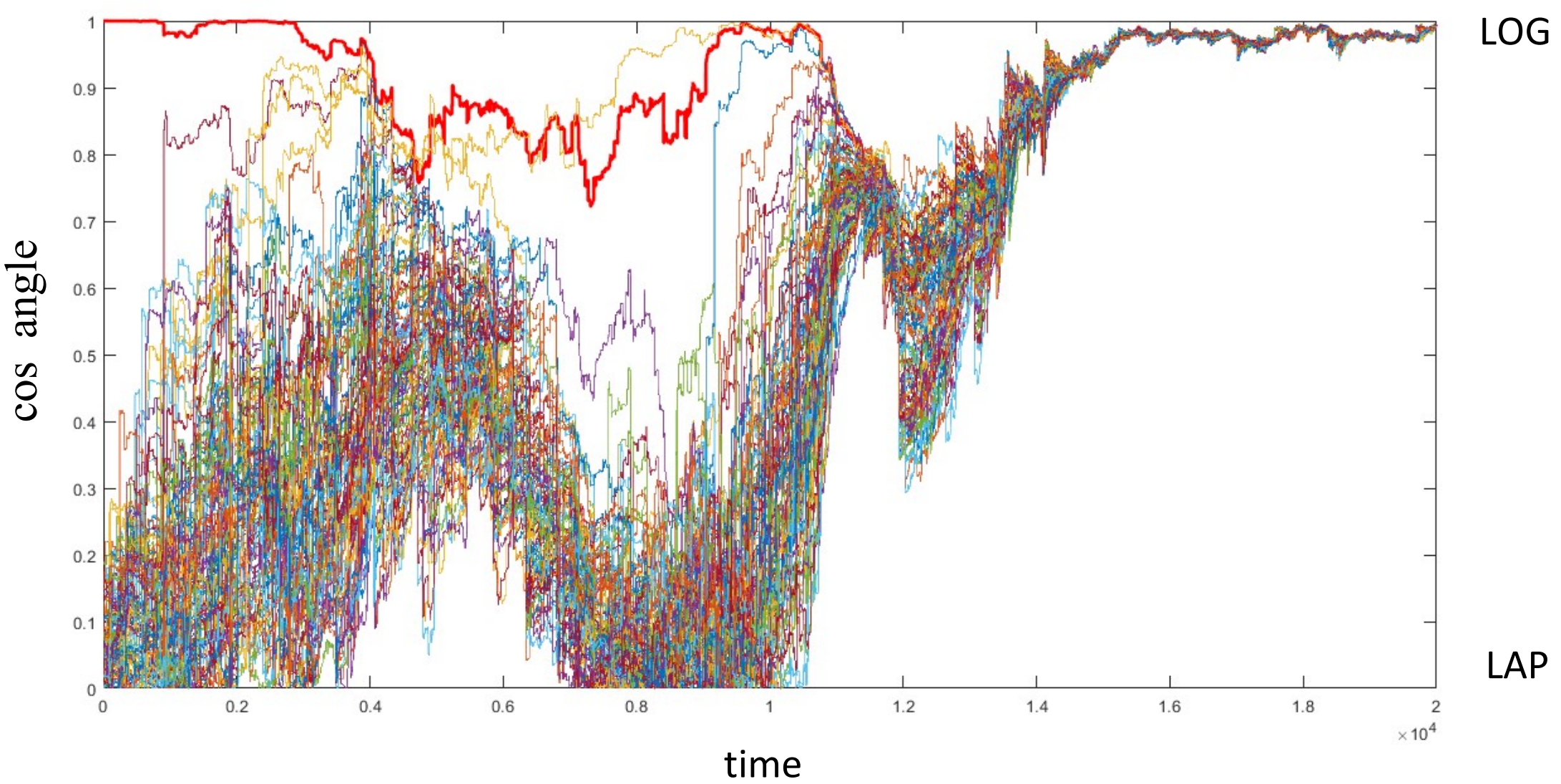

**Figure 9** 100 agents with one starting from (1,-1), i.e. LOG IC with Light (2,5.5). Compare with figure 8. This time all the LAP agents are pulled over to the LOG FP.

## Discussion

Culture can be thought of a set of ideas or (concepts or behaviours) that have been accumulated by a group of 'agents' over time. In the case of humans (where each individual human is an agent) a new concept that an agent discovers is transmitted to other agents that do not yet have that concept. This can be accomplished by teaching, including demonstrations and written explanations. Over at least the past 100,000 years, and especially the last 10,000 years, humans have been able to spread concepts learned by individuals more and more efficiently which has resulted in human culture rapidly becoming more and more 'knowledgeable 'and sophisticated with individuals being able to learn a large number of concepts (learned individually at some point by other agents) which increases their understanding of their environment. However, the way this has occurred is controversial, partly due to the complexity of human society and the problem of defining what a 'concept' is, and in this paper we explore a much simplified model of 'culture' stripped down to just two alternative concepts that can be learned.
In our model there are 2 'concepts' that can be learned from apparently random but nevertheless structured data by an agent, and we can call these concepts the LAP IC and the LOG IC. What we show is that depending on where the agent starts (i.e. the initial weight vector) it will, when learning independently using the ICA learning rule, will move to either the LAP IC or the LOG IC. The LAP basin turned out to be larger than the LOG basin roughly in line with the ratio of kurtoses, and so one could say that the LAP IC was 'easier 'to find.
In this paper we studied a very simple model of the interaction between individual learning (by an agent observing its environment) and social learning (where agents observe, imitate and communicate with each other). The underlying notion is a simple one - individual discovery and invention are slow, difficult, and costly, but learning from others, particularly teachers, who may have already discovered useful solutions, can be quick, easy and advantageous. Thus Newton wrote in a private letter "if I have seen further it's by standing on the shoulders of giants" (Newton, 1960).

Our work is guided by the insight that learning is mainly accomplished by synaptic plasticity (Hayashi-Takagi et al., 2015). Networks of neurons generate complex patterns of electrical activity which map onto behaviors, memories and thoughts, in ways that depend on specific and modifiable synaptic connections. The modifications are triggered by the electrical activity at those connections, in a broadly Hebbian manner (i.e. dependent, inter alia, on conjoint local input and output activity), in such a way that useful behaviors gradually emerge. Hebbian learning is sensitive to correlations, both second- and higher order, and these correlations partially encode causal structure, though extra information (e.g. temporal depedence) is also needed. In previous work we have studied, in highly simplified neural networks, often comprised of single neurons together with their synaptic inputs, how various imperfections (such as inadequate preprocessing or inspecific synaptic modifications, can slow or prevent individual learning. In particular we generated structured observable input patterns by linearly combining statistically independent signals with random but defined distributions. This is probably the simplest possible way to generate structured input patterns that have nontrivial (e.g. higher order) regularities that can be exploited by Hebbian learning to infer (Bishop, 2006; Dayan and Abbott, 2001; Marr, 1982) and the core of intelligence. We studied learning by simplified artificial neurons exposed to these structured patterns using standard learning rules (Hyvärinen and Oja, 1998) slightly modified to be more realistic by incorporating either slight synaptic inspecificity or imperfect preprocessing (Cox and Adams, 2014, 2009) and our conclusions have been reinforced by mathematical analysis.

In an effort to understand how communicating populations of such neurons could surpass individual learning, we extended this model by allowing multiple individual 1-neuron agents to communicate their tentative solutions to others, which could then use more efficient supervised learning to boost or even bypass individual learning. We found that social learning did not help population-level learning, because it accelerates both correct and incorrect individual ;earning, in a manner reminiscent of “Rogers Paradox” in the Cultural Evolution field (Rogers, 1988).

In further work (Cox and Adams, 2023), we showed that this stumbling block could be removed by adding a factor we call “Light”, which selectively enhances learning from individuals that are already close to correct solutions.

Our new results take this one further step. Instead of learning either correct or incorrect solutions, agents learn one of 2 correct solutions, one of which is more difficult to learn than the other. Specifically, because correct learning in the ICA set-up is driven by higher order correlations generated by mixing non-Gaussian signals, the more nonGaussian the source distributions are (as measured by the kurtosis of their distributions), the wider their basin of attraction becomes, most randomly-initialized agents will learn extract Laplacian signals than logistic distributions. If, rather arbitrarily, the Logistic signals are nevertheless more “useful” they can be preferentially learned using the “Light’ factor.

These results are clearly related to the core concepts of Cumulative Cultural Evolution, and may help resolve some of the controversies (Mesoudi et al., 2006) in that relatively recent field of study. The basic idea of CCE is that agents (and specifically humans) can learn to solve difficult survival problems by learning from other agents in a ratchet-like manner. Once a pioneer agent finds a better solution to a survival/reproduction problem than solutions currently prevalent in the population, other agents can socially learn the better solution without the work incurred by the pioneer. The initially prevalent solution would have previously been learned in a similar manner, so that population-level learning becomes cumulative. Previous quantitative studies of CCE (Boyd and Richerson, 1985; Mesoudi et al., 2006; Richerson and Boyd, 2008) have assumed that ‘good’ and ‘better’ solutions are unitary meme-like traits that can be learned as discrete alternative choices, without really specifying how these traits are actually learned, so the ‘inheritance’ and ‘variation’ processes remain mysterious and controversial black boxes. In our model the ‘inheritance’ and ‘variation’ processes are made completely explicit: ‘inheritance’ is done by social, supervised, learning and ‘variation’ is done by individual, unsupervised, learning. Both of these learning processes are quite well understood, both in terms of synaptic/molecular machinery (Hayashi-Takagi et al., 2015), and by detailed theoretical modeling (Bishop, 2006; Dayan and Abbott, 2001). A dramatic example of this is our demonstration (figure 9) that a single ‘pioneer’ agent at a more useful though more difficult to attain solution can coax the entire population to that solution. This hints that progressive cultural shifts to increasingly superior solutions can occur by synaptic learning at a population level.

At least 2 previous studies (Denaro and Parisi, 1997; Gabora, 1995) have proposed Neural Network-based model of Cultural Evolution, with similarities to our own approach. However, these studies use supervised not unsupervised learning from the environment. In the Parisi study networks learn from each other as well as from the environment, and, somewhat similarly to our “Light” proposal, in order for Cultural Evolution to occur, selective learning from agents that have learned good solutions is required. However, innovation comes from noise added to the learning process, rather than from unsupervised learning as in our approach. Furthermore there is no analytic treatment of the problems of individual learning embedded in these papers.

It is our view (Cox and Adams, 2021) that unsupervised learning is more fragile than supervised learning, in the sense that an additional supervised element can overcome difficulties such as imperfect whitening, synaptic crosstalk, false minima, and flat regions in weight space. Nevertheless unsupervised learning is likely to play a central role in discovery (Hinton and Sejnowski, 1999). And in a sense this is what Cultural Evolution is all about: converting unsupervised learning (from an unlabeled environment) into a much easier supervised task (learning from other individuals who have learned, directly or indirectly, without guiding labels). Supervision essentially uses unsupervised routes, ropes and pitons placed by pioneering climbers.

In particular our results suggest that the key condition under which CCE can occur is not high-fidelity social learning (Boyd and Richerson, 1985; Mesoudi et al., 2006), but instead a combination of 2 factors: synergistic interaction between social and individual learning, and selective boosting of social learning by light-like factors. Our mathematical formulation of 'Light' is deliberately vague: it could correspond to "prestige", "trust", or even "science". In our model agents learn continuously and incrementally, alternating between supervised observation of other agents and direct unsupervised learning from a shared environment. High-fidelity learning, both supervised and unsupervised, helps of course, and indeed we have explicitly modeled the effects of low-fidelity learning using an "error-matrix' description, but in our models it is not a helpful source of 'variation". Ultimately learning a complex concept or behavior must involve the slow progressive adjustment of enormous numbers of synaptic "weights" (Lisberger, 2024; Markham and Greenough, 2004), in a manner that's suggested by the various plots in our "Results" section, and during the population-level learning process there's a complex dance between different interacting partners' weights, rather than a simple direct "copying" process. Cognitive and behavioral "traits" are useful simplifications of the detailed neural processes underlying learning.

However, we acknowledge that the simplified setup we use here, with only 2 stable fixed points and dynamics on the unit circle, with 2 distinct basins of attraction, might be somewhat misleading. If there were 3 or more fixed points the dynamics become much richer. For example, if we added crosstalk to our model, the learning dynamics would be on the unit sphere, and in addition to stable and unstable fixed points there could also be saddle points. In the current model, one could regard basins of attraction as discrete cultural traits or "memes", as in conventional CE theory. But on the sphere, or a fortiori the hypersphere, weight configurations that are close to saddle points will emerge. These configurations would be ambiguous, and this ambiguity would not be fully captured in traditional CE terminology (idea fragments and recombination). We are currently exploring these interesting possibilities. In particular, it's possible that in these circumstances populations of interacting agents could learn synaptic weight configurations that are inaccessible to isolated, noninteracting individual. We are currently exploring this "collective intelligence" possibility.

A possible weakness of our approach is the simplicity of the ICA framework itself. However, we believe that this framework, and the associated mathematical transparency, allows clearer insight into the hidden synaptic machinery implicit in traditional CE models. The core issue is that the brain's task, making sense of an apparently random world, is extremely difficult, and that cooperation between agents could overcome some of these difficulties. This issue is already at play in our simplified ICA framework (Hyvärinen et al., 2018) (hyv). Furthermore, recent progress in nonlinear ICA shows how simple ICA might be a stepping stone to general intelligence. And there are intriguing hints that early visual cortex might perform ICA-like operations (Bell and Sejnowski, 1997).

# Appendix

**Basin Structure in One-Unit ICA with Cubic Nonlinearity**

We consider a one-unit independent component analysis (ICA) model with two independent, zero-mean, unit-variance sources $s_1$ and $s_2$. After whitening, the observed data vector is given by:

$$x = \binom{s_1}{s_2}$$

The weight vector $w$ is constrained to unit norm and can therefore be parameterised by a single angle $\theta$:

$$x = \binom{cos\theta}{sin\theta}$$

The scalar output of the unit is:

$$y = \mathbf{w}^T\mathbf{x} = cos\theta s_1 + sin\theta s_2$$

**Objective Function**

For a cubic nonlinearity $g(y) = y^3$, the associated contrast function is proportional to the fourth moment:

$$J(\theta) = \mathbb{E}[y^4]$$

Expanding $y^4$ and using independence and symmetry of the sources yields:

$$\mathbb{E}[y^4] = cos^4\theta\ \mathbb{E}[s_1^4] + sin^4\theta\ \mathbb{E}[s_2^4] + 6cos^2\theta cos^2\theta$$

Introducing the excess kurtosis $\kappa_i = \mathbb{E}\left[s_i^4\right] - 3$, the objective can be written (up to an additive constant) as:

$$J(\theta) = \kappa_1 cos^2\theta + \kappa_2 sin^2\theta + const$$

## Fixed Points and Separatrix

Differentiating with respect to $\theta$ gives:

$$\frac{\mathrm{dJ}}{\mathrm{d\theta}} = 4\cos\theta\sin\theta(\kappa_2\sin^2\theta - \kappa_1\cos^2\theta)$$

Setting $\frac{\mathrm{dJ}}{\mathrm{d\theta}}=0$ yields fixed points at $\theta = 0$ and $\theta = \frac{\pi}{2}$, corresponding to alignment with the independent components. In addition, there is an interior stationary point satisfying:

$$\kappa_2\sin^2\theta - \kappa_1\cos^2\theta$$

which gives:

$$\tan^2\theta_* = \frac{\kappa_1}{\kappa_2}$$

This interior point is a saddle and defines the separatrix between the basins of attraction.

**Basin Sizes**

The dynamics are confined to the unit circle, so $\theta \in \left[0, \frac{\pi}{2}\right]$. The separatrix is located at:

$$\theta_* = \left(\sqrt{\frac{\kappa_1}{\kappa_2}}\right)$$

Thus, the basin of attraction of source $s_1$corresponds to $0 \leq \theta < \theta_*$, while that of $s_1$corresponds to $\theta_* \leq 0 < \frac{\pi}{2_*}$ . The relative basin sizes are therefore:

$$Basin\, ratio = \frac{\theta_*}{\frac{\pi}{2} - \theta_*}$$

**Example**

For unit-variance sources, the Laplace distribution has excess kurtosis $\kappa_{Lap} = 3$, while the logistic distribution has $\kappa_{Lap} = 1.2$. Substituting these values gives:

$$\theta_* = \frac{3}{1.2} = 2.5$$

$$\theta_* = \arctan\left(\sqrt{2.5}\right) \approx 57^\circ$$

This yields basin sizes of approximately $57^\circ$and $33^\circ$, giving a ratio of:

$$\frac{57}{33} \approx 5:3$$

---

**Interpretation**

The basin structure is determined by the ratio of source kurtoses. Higher-kurtosis sources possess larger basins of attraction, implying that random initialisations are more likely to converge to those components. This establishes a direct quantitative relationship between higher-order statistics and recovery probability in one-unit ICA.